\newcommand{\diracslash}[1]{#1\llap{/\kern2pt}}

\newcommand{\be}{\begin{equation}}
\newcommand{\ee}{\end{equation}}
\newcommand{\bea}{\begin{eqnarray}}
\newcommand{\eea}{\end{eqnarray}}
\newcommand{\ba}[1]{\begin{array}{#1}}
\newcommand{\ea}{\end{array}}

\documentclass[prl,aps,floats,nofootinbib,tightenlines,showpacs,twocolumn,floatfix]{revtex4}
\usepackage{graphicx}
\begin{document}
\title {Kinetics of chiral phase transition in hot and dense quark matter} 
\author{Awaneesh Singh }
\email{awaneesh11@gmail.com}
\affiliation{School of Physical Sciences, Jawaharlal Nehru University,
New Delhi 110067, India}
\author{Sanjay Puri}
\email{purijnu@gmail.com}
\affiliation{School of Physical Sciences, Jawaharlal Nehru University,
New Delhi 110067, India}
\author{Hiranmaya Mishra}
\email{hm@prl.res.in}
\affiliation{School of Physical Sciences, Jawaharlal Nehru University,
New Delhi 110067, India}
\affiliation{Theory Division, Physical Research Laboratory,
Navrangpura, Ahmedabad 380 009, India}

\date{\today} 

\def\be{\begin{equation}}
\def\ee{\end{equation}}
\def\bearr{\begin{eqnarray}}
\def\eearr{\end{eqnarray}}
\def\zbf#1{{\bf {#1}}}
\def\bfm#1{\mbox{\boldmath $#1$}}
\def\hf{\frac{1}{2}}
\begin{abstract}
The kinetics of chiral transitions in hot and dense quark matter is studied via a microscopic framework (Nambu-Jona-Lasinio model) and a phenomenological model (Ginzburg-Landau free energy). We focus on the far-from-equilibrium ordering dynamics subsequent to a quench from the massless quark phase to the massive quark phase. The morphology of the ordering system is characterized by the scaling of the order-parameter correlation function. The domain growth process obeys the Allen-Cahn growth law, $L(t)\sim t^{1/2}$. We also study the growth of bubble of the stable massive phase in a background of the metastable massive phase.
\end{abstract}

\pacs{12.38.Mh, 24.85.+p} 

\maketitle
The kinetics of phase transitions, and the ordering process that occurs after a rapid quench in system parameters like temperature and pressure, plays an important and interesting role in different areas of physics \cite{pw09}. During the transition, the system develops a spatial structure of randomly-distributed domains which evolve with time. This ordering process has been extensively studied in many condensed matter systems like ferromagnets, binary fluids, liquid crystals, etc. In this letter, we study an equally fascinating application in high-energy physics, i.e., the quark hadron phase transition. 

\par Heavy ion collision experiments at high energies produce hot and dense strongly-interacting matter, and provide the opportunity to explore the phase diagram of QCD in the plane of temperature ($T$) and baryon chemical potential ($\mu$). Many model studies \cite{mk98} as well as recent lattice studies \cite{latmu} indicate that at sufficiently large baryonic densities, there is a line of first-order transitions in the ($\mu, T$)-plane between a chirally-symmetric phase and a broken-symmetric phase. As one moves along the phase boundary towards higher $T$ and smaller $\mu$, the first-order transition becomes weaker-- ending in a second-order critical point in the limit of vanishing current quark mass or a rapid crossover for nonzero current quark mass \cite{misha}. While the high-$T$  and small-$\mu$ region of the QCD phase diagram has been explored in recent experiments, future heavy-ion  collision experiments plan to explore the high baryon density regime, particularly the region around the critical point \cite{cpod}. 

\par In this context, the critical dynamics of the chiral transition, and its static and dynamical universality properties, have attracted much attention recently. Signatures of the critical point which had been proposed using on equilibrium thermodynamics \cite{shuryakprl}, have been recently interpreted within a quasi-stationary framework \cite{misha2}. In the present letter, we study the far-from-equilibrium kinetics of chiral phase transitions subsequent to a quench from a disordered phase (with vanishing quark condensate) to the ordered phase. 

\par
\begin{figure}[!hb]
\vspace{-0.44cm}
\centering
\includegraphics[width=0.19\textwidth]{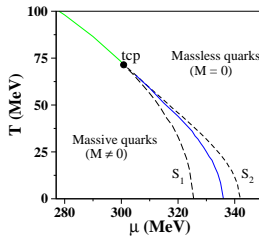}
\vspace{-0.4cm}
\caption{Phase diagram of the NJL model in the ($\mu, T$)-plane for zero current quark mass. A line of first-order transitions (green online) meets a line of second-order transitions (blue online) at the tricritical point (tcp). We have $(\mu_{tcp}, T_{tcp}) \simeq (285.0, 74.9)$ MeV. The dashed lines denote the spinodals for the first-order transitions.}
\label{fig1}
\end{figure}

 To model chiral symmetry breaking in QCD, we use the two-flavor Nambu-Jona-Lasinio (NJL) model \cite{klevansky} with the Hamiltonian:
\begin{eqnarray}
{\cal H} &=& \sum_{i,a}\psi^{ia \dagger}\left(-i{\mbox{\boldmath {$\alpha$}}}\cdot{\mbox{\boldmath {$\nabla$}}} + \gamma^0 m_i \right)\psi^{ia} \nonumber \\
&& \qquad -G\left[(\bar\psi\psi)^2-(\bar\psi\gamma^5 {\mbox{\boldmath {$\tau$}}} \psi)^2\right].
\label{ham}
\end{eqnarray}
To describe the ground state, we take an ansatz with quark-antiquark condensates \cite{hmnj}:
\begin{equation} 
|vac\rangle= \exp\!\!\left[\!\int\!\! d\zbf k~ q_I^{0i}(\zbf k)^\dagger(\mbox{\boldmath {$\sigma$}}\cdot\zbf k)h_i(\zbf k)\tilde q_I^{0i} (-\zbf k)-\mathrm{h.c.}\right]\!|0\rangle.
\label{uq}
\end{equation}
Here, $q^\dagger$, $\tilde q$ are two-component quark and antiquark creation operators, and $|0\rangle$ is the perturbative chiral vacuum. Further, $h_i(\zbf k)$ is a variational function related to the quark-antiquark condensate as $\langle\bar{\psi}\psi\rangle = -6/(2\pi)^3 \sum_{i=1}^2 \int \!d\zbf k\,\sin[2h_i(\zbf k)]$. This flavor-dependent function can be determined by minimising the energy at $T=0$ or the thermodynamic potential at nonzero $T$ and density. Without going into the details which are reported elsewhere \cite{awaneesh1}, we write down the expression for the thermodynamic potential as 
\begin{eqnarray}
\tilde\Omega(M,\beta,\mu) &=&-\frac{12}{(2\pi)^3\beta}\sum_{i=\pm}\int d\zbf k~
\{\ln\left[1+\exp(-\beta\omega_i)\right]\}  \nonumber\\
&&-\frac{12}{(2\pi)^3}\int d\zbf k~ \left(\sqrt{\zbf k^2+M^2}-|\zbf k|\right)\nonumber\\
&&+g\rho_s^2 -\frac{G}{2 N_c}\rho_v^2.
\label{tomega}
\end{eqnarray}
Here, we have taken vanishing current quark mass, and introduce $g=G[1+1/(4N_c)]$; $M=-2g\rho_s$ with $\rho_s=\langle\bar\psi\psi\rangle$ being the scalar density. Further, $\rho_v=\langle\psi^\dagger\psi\rangle$ is the vector density, and $\omega_\pm= (\zbf k^2+M^2)^{1/2}\pm\nu$ where $\nu= \mu-G\rho_v/N_c$. The resulting phase diagram is shown in Fig. \ref{fig1}. We have taken here a three-momentum ultraviolet cutoff $\Lambda=631$ MeV, and the four-fermion coupling $G=5.074\times 10^{-6}$ $\mathrm{MeV^{-2}}$ \cite{askawa}. With these parameters, the vacuum mass of quarks is $M\simeq 321$ MeV. At $T=0$ the first-order transition takes place at $\mu \simeq 335$ MeV. For $\mu=0$, a second-order transition takes place at $T \simeq 195$ MeV. The first-order line meets the second-order line at the tri-critical  point $(T_{tcp},\mu_{tcp})\simeq(74.9,285.0)$ MeV. The first-order transition is characterized by the existence of metastable phases. The limit of metastability is denoted by the dashed lines in Fig. \ref{fig1}, referred to as the spinodal lines.
\par Close to the phase boundary, the thermodynamic potential may be expanded in a power series of the order parameter $M$: 
\be
\tilde\Omega\left(M \right)= \tilde\Omega\left(0 \right) + \frac{a}{2}M^2 + \frac{b}{4}M^4 + \frac{d}{6}M^6 + \cdots \equiv F\left(M \right).
\label{p6}
\ee
One can obtain the coefficients in terms of $\beta$ and $\mu$  by Taylor-expanding the thermodynamic potential in Eq. (\ref{tomega}). However, we will treat them as phenomenological constants which are adjusted to recover the phase diagram in Fig. \ref{fig1}. After all, we are interested in studying the kinetics of the chiral transition, usually at parameter values far from the critical curve in Fig. \ref{fig1}-- where we cannot Taylor-expand $\tilde\Omega\left(M,\beta,\mu \right)$. The extrema of the potential in Eq. (\ref{p6}) are determined by the gap equation $F'(M)=aM+bM^3+dM^5=0$. The solutions are $ M_0=0$, and $M_{\pm}^2=(-b\pm \sqrt{b^2 -4ad})/{2d}$. For $b>0$, the transition is second-order, analogous to an $M^4$-potential-- the stationary points are $M=0$ (for $a>0$) or $M=0$, $\pm M_+$ (for $a<0$). For $a<0$, the preferred equilibrium state is the one with massive quarks. For $b<0$, the solutions of the gap equation are as follows: (i) $M=0$ for $a>b^2/4d$, (ii) $M=0$, $\pm M_+$, $\pm M_-$ for $b^2/4d>a>0$, and (iii) $M=0$, $\pm M_+$ for $a<0$. A first-order transition takes place at $a_c=3b^2/16$ with the order parameter jumping discontinuously from $M=0$  to $M=\pm M_+=\pm (3|b|/4d)^{1/2}$. The tricritical point is located at $b_{tcp}=0$, $a_{tcp}=0$.

\par Let us consider the dynamical environment of a heavy-ion collision. As long as the evolution is slow compared to the typical re-equilibration time, the order parameter field will be in local equilibrium. We consider a system which is rendered thermodynamically unstable by a rapid quench from the disordered phase to the ordered phase. The unstable disordered state evolves via the emergence and growth of domains rich in the preferred phase \cite{pw09, aj94}. The coarsening system is inhomogeneous, and we include a surface tension term in the Ginzburg-Landau free energy as follows:
\begin{equation}
\Omega(M)=\int d\zbf r \left[\frac{a}{2} M^2+\frac{b}{4} M^4+\frac{d}{6} M^6
+\frac{K}{2}\left(\zbf\nabla M\right)^2\right].
\label{omgl}
\end{equation}
The evolution of the system is described by the time-dependent Ginzburg-Landau (TDGL) equation 
\begin{equation}
\frac{\partial}{\partial t}M\left(\zbf{r},t\right)= -\Gamma 
\frac{\delta \Omega\left[M \right]}{ \delta M}+\theta\left(\zbf{r},t\right), 
\label{ke}
\end{equation}
modeling the over-damped relaxational dynamics of $M(\zbf r,t)$ to the minimum of $\Omega\left[M \right]$ \cite{hohenrev}. Here, $\Gamma$ is the inverse damping coefficient, and $\theta(\zbf r,t)$ is the noise term satisfying the fluctuation-dissipation relations $\left\langle\theta\left(\zbf{r},t\right) \right\rangle = 0$ and $\left\langle \theta(\zbf{r'},t')\theta(\zbf{r''},t'') \right\rangle = 2T\Gamma\delta(\zbf{r'}-\zbf{r''})\delta\left(t'-t''\right)$. We use the natural scales of order parameter, space and time to introduce the dimensionless variables, $M=\sqrt{|a|/|b|}~M'$, $\zbf r=\sqrt{K/|a|}~\zbf r'$, $t=t'/(\Gamma|a|)$, $\theta=(\Gamma|a|^{3/2}/|b|^{1/2})~\theta'$. Dropping the primes, we have the dimensionless TDGL equation:
\begin{eqnarray}
\frac{\partial}{\partial t}M\left(\zbf{r},t\right)&=& -\mathrm{sgn}
\left(a\right)M - \mathrm{sgn}\left(b\right)M^3 \nonumber \\
&& -\lambda M^5 + \nabla^2 M +\theta\left(\zbf{r},t\right),
\label{ke2}
\end{eqnarray}
where $\lambda=|a|d/b^2$. For $T=15$ MeV, as $\mu$ takes values 320 MeV, 330 MeV, 338 MeV and 350 MeV, the corresponding values of $\lambda$ are 0.065, 0.14, 0.23 and 0.34, which we have obtained by fitting the effective potential Eq. (\ref{tomega}) with Eq. (\ref{p6}).

\par First, we study the ordering dynamics for $b>0$. We solve the Eq. (\ref{ke2}) numerically using an Euler-discretization scheme. This is implemented on a 3-d lattice of size $N^3$ ($N=256$), with periodic boundary conditions in all directions. The dimensionless mesh sizes are $\Delta x=1.0$ and $\Delta t=0.1$, which satisfy the numerical stability condition. We have further confirmed that the spatial mesh size is sufficiently small to resolve the interface region. The details of the numerics will be reported elsewhere \cite{awaneesh1}. 
\begin{figure}[!htb]
\vspace{-0.6cm}
\centering
\includegraphics[width=0.46\textwidth]{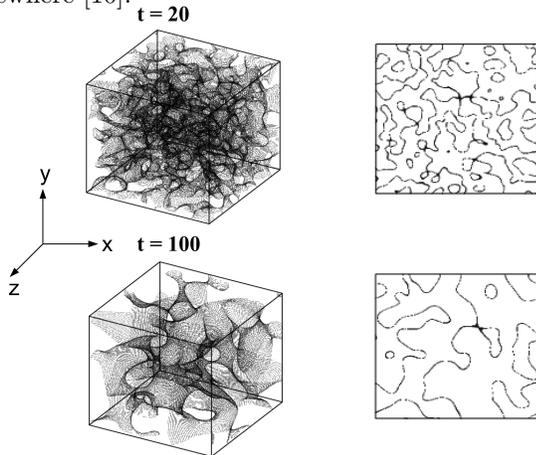}
\vspace{-0.48cm}
\caption{Interface evolution after temperature quench through second-order line in Fig. \ref{fig1}. The 3-d snapshots on the left show the interfaces ($M=0$) at $t=20$, $100$. They were obtained by numerically solving Eq. (\ref{ke2}) with $a<0$, $b>0$, $\lambda=0.14$. The noise amplitude was $\epsilon=0.008$. The frames on the right show a cross-section of the snapshots at $z=N/2$.}
\label{fig2}
\end{figure}
\begin{figure}[!htb]
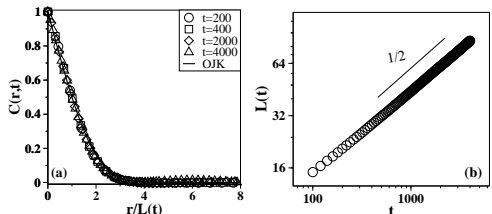

\vspace{-0.5cm}
\begin{center}
\begin{tabular}{c c }
\includegraphics[width=0.175\textwidth]{fig3a.eps}&
\includegraphics[width=0.175\textwidth]{fig3b.eps}\\
\vspace{-1.1cm}
\end{tabular}
\end{center}
\caption{(a) Shows the scaling of correlation function for $\lambda=0.14$ at four different time steps. OJK function (as for usual $M^4$-free energy) has good agreement with simulation data. (b) The domain size $L(t)$ vs $t$ for $\lambda=0.14$. Domain growth data is consistent with Allen-Cahn growth law $L(t)\sim t^{1/2}$}
\label{fig3}
\end{figure}
\par In Fig. \ref{fig2}, we show the evolution of a disordered initial condition for Eq. (\ref{ke2}) with $a<0$, i.e., a temperature quench through the second-order line in Fig. \ref{fig1}. The initial state consisted of small-amplitude thermal fluctuations about the massless phase $M=0$. The system rapidly evolves into domains of the massive phase with $M\simeq M_+$ and $M\simeq-M_+$. The interfaces correspond to $M=0$-- their evolution is shown in the snapshots (frames on left) of Fig. \ref{fig2}. The frames on the right shows the interface structure in a cross-section of the snapshots. The domains have a characteristic length scale which grows with time. The growth 
process is analogous to the coarsening dynamics in the TDGL equation with an $M^4$-potential, where growth is driven by kinks with the equilibrium profile $M(z)=\tanh(\pm z/\sqrt 2)$. 
The order-parameter correlation function $C(r,t)$ shows \emph{dynamical scaling} or \emph{dynamical self-similarity}, $C(r,t)=f(r/L)$-- the scaling function $f(x)$ has been calculated by Ohta et al. (OJK) \cite{ojk82}. In Fig. \ref{fig3}(a), we demonstrate that $C(r,t)$ for the evolution in Fig. \ref{fig2} shows the scaling property. Further, the domain scale obeys the Allen-Cahn (AC) growth law, $L(t)\sim t^{1/2}$ [see Fig. \ref{fig3}(b)]. The interface velocity $v\sim dL/dt \sim 1/L$, where $L^{-1}$ measure the local curvature of the interface. This yields the AC growth law.

\par
\begin{figure}[!htb]
\vspace{-0.3cm}
\centering
\includegraphics[width=0.35\textwidth]{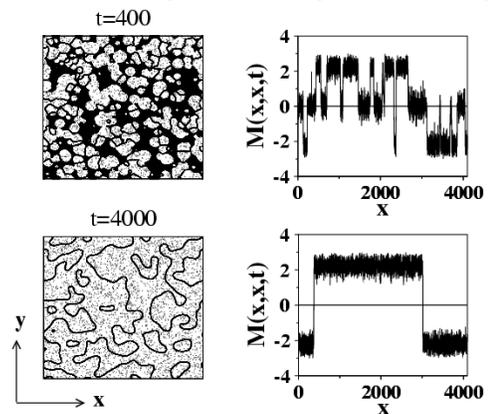}
\vspace{-0.38cm}
\caption{Quenching through first-order phase transition line. Evolution of regions with $M=0$ (black colored) for $\lambda=0.14$ and for times $t=400$ and $t=4000$. To the right of the each evolution pattern, the order parameter profiles along the diagonal are also shown.}
\label{fig4}
\end{figure}

 Next, let us consider the case with $b<0$. In this case, a first-order chiral transition occurs for $a<a_c=3|b|^2/(16d)$ (or $\lambda<\lambda_c=3/16)$. For $a<0$, the potential has a double well structure and the ordering dynamics is analogous to $M^4$-theory i.e., the domain growth scenario is similar to Fig. \ref{fig3}. We focus our attention on a quench from the disordered state ($M=0$) to  $0<\lambda< \lambda_c$, corresponding to a quench between the first-order line and $\mathrm{S_1}$ in Fig. \ref{fig1}. The massless state is now a metastable state of the $M^6$-potential. The chiral transition proceeds via the nucleation and growth of droplets of the preferred phase ($M=\pm M_+$). This nucleation results from large fluctuations in the initial condition or thermal fluctuations during the evolution. In Fig. \ref{fig4}, we show the nucleation and growth process. At early times ($t=400$) the system is covered with the $M=0$ phase with small droplets of the preferred phase. These droplets grow in time and coalesce into domains. The subsequent coarsening of these domains is analogous to that in Figs. \ref{fig2}-\ref{fig3}. In the late stages, there is no memory of the nucleation dynamics which characterized growth during the early stages.
\begin{figure}[!htb]
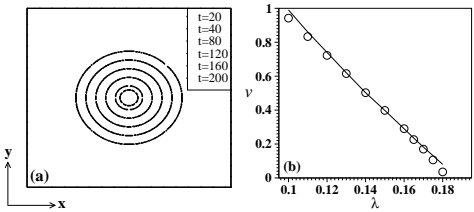

\vspace{-0.4cm}
\begin{center}
\begin{tabular}{c c }
\includegraphics[width=0.17\textwidth]{fig5a.eps}&
\includegraphics[width=0.17\textwidth]{fig5b.eps}\\
\vspace{-1.1cm}
\end{tabular}
\end{center}
\caption{(a) Shows the growth of single droplet of the preferred phase ($M = +M_+$) in the background of the metastable phase ($M=0$) for $\lambda = 0.14$ for different time steps. The inner circle corresponds to a droplet at time, $t=20$. (b) Shows the plot of droplet velocity $v$ vs. $\lambda$. Circles refer to numerical data while solid line corresponds to predictions from a phase-plane analysis.}
\label{fig5}
\end{figure}
\par We have also studied the growth of single droplets of the preferred phase ($M=+M_+$) in the background of the metastable phase ($M=0$) [see Fig. \ref{fig5}(a)]. The droplets have a unique growth velocity $v(\lambda)$, which depends on the degree of undercooling ($\lambda$). We have obtained $v(\lambda)$ by undertaking a phase-plane analysis of the travelling-wave solutions of Eq. (\ref{ke2}). The droplet interface corresponds to a saddle connection between the fixed points $+M_+ \rightarrow 0$ or $-M_+ \rightarrow 0$. The details of this analysis will be presented elsewhere \cite{awaneesh1}. In Fig. \ref{fig5}(b), we plot numerical results for $v(\lambda)$ vs. $\lambda$ along with our theoretical prediction. Before concluding, we briefly discuss the inertial kinetic equation, which is the counterpart of the overdamped  TDGL equation (\ref{ke}). The TDGL equation contains first-order time-derivatives and is not Lorentz invariant. This is not a serious problem as we are considering finite-temperature field theory, where there is no Lorentz invariance.  Nevertheless, one can also study the kinetics of chiral transitions via an inertial evolution equation:
\begin{eqnarray}
\left(\frac{\partial^2}{\partial t^2}-\zbf{\nabla^2} \right)M 
+ \gamma\frac{\partial}{\partial t}M &=&
 -\mathrm{sgn}\left(a\right)M - \mathrm{sgn}\left(b\right)M^3  \nonumber \\
&& - \lambda M^5  +\theta\left(\zbf{r},t\right).
\label{ke1}
\end{eqnarray}
  
\par In principle, Eq. (\ref{ke1}) can be obtained from a microscopic field-theoretic description of the nonequilibrium dynamics of the scalar field at finite temperatures \cite{ramos}. We have also studied the ordering dynamics for the inertial case, and will present details of our simulations elsewhere \cite{awaneesh2}. Here, we mention the main results  of our study. The ordering dynamics in this case is similar to that in the overdamped case, except that nucleation does not have a significant effect even during the early stages of evolution. The droplets grow very rapidly and merge to form a bicontinueous domain structure characteristic of late stage domain growth. The domain growth law is again the AC law $L(t)\sim t^{1/2}$.

\par To summarize, we have studied the kinetics of the chiral phase transition in QCD. In terms of the quark degrees of freedom, the equilibrium phase diagram is obtained using the NJL model. An equivalent coarse-grained description is obtained from an $M^6$-Ginzburg-Landau (GL) free energy. We study the chiral kinetics via the corresponding TDGL equation, and consider both the overdamped and inertial cases. We study the ordering dynamics resulting from a sudden quench in temperature through the first-order (I) or second-order (II) transition lines in Fig. \ref{fig1}. For quenches through II and deep quenches through I, the massless phase is spontaneously unstable and evolves to the ordered phase via spinodal decomposition. For shallow quenches through I, the massless phase is metastable and the phase transition proceeds via nucleation and growth of droplets. The subsequent merger of these droplets results in late-stage domain growth analogous to that for the unstable case. In all cases, the late-stage growth process exhibits self-similar dynamical scaling, and the growth law is $L(t)\sim t^{1/2}$.
\vspace{0.5cm}\\
{\bf Acknowledgment}\\
AS would like to thank CSIR (India) for financial support. HM would like to thank the School of Physical Sciences, Jawaharlal Nehru University, New Delhi for hospitality.

\end{document}